\documentclass[twocolumn,aps,floats,superscriptaddress]{revtex4}
\usepackage{amsmath}

\usepackage{epsfig}
\usepackage{graphicx}
\usepackage{dcolumn}
\usepackage{bm}
\begin{document}
%
\title{
Electron-Phonon Interacation in Quantum Dots: \protect\\
A Solvable Model
}
\author{T. Stauber}
\affiliation{Institut f\"ur Physik, Humboldt-Universit\"at zu Berlin,
Hausvogteiplatz 5-7, D-10117 Berlin, Germany}
\author{R. Zimmermann}
\affiliation{Institut f\"ur Physik, Humboldt-Universit\"at zu Berlin,
Hausvogteiplatz 5-7, D-10117 Berlin, Germany}
\author{H. Castella}
\affiliation{Max-Planck-Institut f\"ur Physik komplexer Systeme,
N\"othnitzer Str. 38, D-01187 Dresden, Germany}

\date{\today}
\begin{abstract}
The relaxation of electrons in quantum dots via phonon emission is
hindered by the discrete nature of the dot levels (phonon bottleneck).
In order to clarify the issue theoretically we consider a
system of $N$ discrete fermionic states (dot levels) coupled to
an unlimited number of bosonic modes with the same energy (dispersionless
phonons). In analogy to the Gram-Schmidt orthogonalization
procedure, we perform a unitary transformation into new bosonic modes.
Since only $N(N+1)/2$ of them couple to the fermions, a numerically
exact treatment is possible.
The formalism is applied to a GaAs quantum dot with only two electronic
levels. If close to resonance with the phonon energy, the electronic
transition shows a splitting due to quantum mechanical level repulsion.
This is driven mainly by one bosonic mode, whereas the other two
provide further polaronic renormalizations.
The numerically exact results for the electron spectral function compare favourably with an analytic solution based on degenerate
perturbation theory in the basis of shifted oscillator states.
In contrast, the widely used selfconsistent first-order Born approximation
proves insufficient in describing the rich spectral features.
\end{abstract}
%
\pacs{71.38.+i; 73.61.Ey}
\maketitle
%
\section{Introduction}
%
Since the development of quantum well lasers there have been continuous
attempts to manufacture laser structures with even more reduced dimensions.
The idea behind was to increase the efficiency by enhancing the density
of states. However, zero-dimensional quantum structures (quantum dots)
are characterized by a discrete spectrum, and the recombination probability
does not depend on the radiative rate alone. Rather, the relaxation pathway
into the groundstate becomes decisive. Looking at the nearly monoenergetic
longitudinal-optical (LO) phonons, an efficient relaxation between two dot
levels seems to be possible only if level distance and LO energy match
(resonance condition).
This type of argument has been called phonon bottleneck.
\cite{BockBast,Beni}

>From the experimental side there is an ongoing intense debate
on whether or not the phonon bottleneck is seen in the data.\cite{Murdin}
However, the recently found groundstate lasing in quantum dots
under cw conditions seems to prove that the phonon bottleneck
is not an obstacle when trying to increase the laser efficiency by
dimensional reduction.\cite{DotLaser}

Nevertheless the theoretical concepts are still controversial. Obviously,
the bottleneck argument relies on the assumption of strict energy
conservation in the electron-phonon scattering, as dictated by Fermi's
golden rule. A next step towards a realistic description seems to
incorporate the intrinsic lifetime broadening of dot levels.
Kr\'al {\it et al.}\cite{KralKhas} went along
this way by calculating the complex electron selfenergy due to
LO-phonon interaction. They claimed that the convolution of initial and
final state spectral functions gives rise to a broadening which is able
to circumvent the phonon bottleneck. Arakawa and coworkers\cite{LiNakaAra,LiAra}
have treated the electronic transition and the LO modes in closed form by
wave function evolution. They pointed out that the final decay of the
LO phonon into acoustic phonons is decisive for the relaxation process.
A different argument includes Auger-like electronic excitation for overcoming
the sharp energy selection inherent to the LO-phonon
relaxation.\cite{Bockel,FerBast,Guy}

A more general question concerns the appropriate theoretical tools
for describing relaxation in zero-dimensional systems.
Non-equilibrium Green's functions are often too tedious to be used
in realistic models. Therefore one is tempted to look for the
one-particle Green's functions and their broadening as a signature for
relaxation.\cite{KralKhas,InoSaka} However, standard selfenergy approaches
as the selfconsistent first-order Born approximation \cite{KralKhas} have
to be questioned since they rely implicitly on the existence of an
electronic continuum which is missing in zerodimensional systems.

In this paper we want to look closely at this question and to qualify
the standard approximation schemes (as the selfconsistent Born approximation) in
application to quantum dots. To compare with we present results from an
exact diagonalization of the electron-phonon Hamiltonian. This can be
achieved even for an unlimited number of phonon modes provided they have
{\it no dispersion},
and uses a unitary transformation among the phonons. If $N$ electronic
dot states (fermions) are considered, only $N(N+1)/2$ of the new modes
(bosons) couple to the electrons, and for moderate numbers of $N$ the
transformed Hamiltonian can be easily diagonalized numerically. An upper
limit of the boson occupation numbers can be fixed in accordance with
temperature. For $N=1$, this
exact solution is known for a long time as {\it independent Boson
model}.\cite{Mahan} To the best of our knowledge, the extension to a finite
number of levels with the important interlevel coupling (phonon transition)
is presented here for the first time.
A {\it single} boson mode in resonance with an
equidistant series of electronic levels has been treated exactly in
Ref.\ \onlinecite{Kenrow} predicting the phonon staircase effect.

A related problem is the electron-phonon coupling in semiconductor point
defects. Both the internal defect transitions in the near infrared and
the phonon satellites of interband transitions show a rich spectrum. Even
away from strict resonance with an electronic transition, the measured
transition energies differ from the bare phonon value. This has been
called LO phonon-donor bound state, similar to the exciton-phonon complex
introduced earlier.\cite{Toyozawa} In first attempts for a quantitative
understanding, perturbation theory has been used.\cite{Dean}
The crystallographic symmetry of the defect dictates what kind of
lattice distortion (local phonon mode) couples to the electronic
transitions. Taking into account only a few of these symmetry-adapted
lattice modes, a full numerical diagonalization is possible
nowadays.\cite{Savona} This is in particular important when dealing with
strongly polar material.
We show that semiconductor quantum dots behave similarly with respect to
the lattice coupling, underlining once more that the quantum dot behaves
as a kind of mesoscopic atom. Note, however, the quite different length
scales involved. Whereas the local modes in the defect problem
are constructed using large parts of the Brillouin zone, only a minor
part around the $\Gamma$ point is involved in the quantum dot case.
Consequently, the LO phonon dispersion can be safely neglected here.

In Section II the method is outlined and applied to the most simple but
nontrivial case of two dot levels. Results for the spectral function
are given in Section III. It shows a kind of avoided level crossing if
level distance and phonon energy nearly coincide, which resembles the
phonon polariton feature. Still, the exact spectral function consists of
a series of sharp lines. The selfconsistent first order Born approximation
fails in this respect by exhibiting broad spectral features.
However, we are able to derive a simple analytical approximation which
almost coincides with the exact results. This employs degenerate
perturbation theory for those electron-phonon states which are strongly
coupled near resonance. It is called rotating wave approximation since
it sesemble a similar treatment of the Jaynes-Cummings model in
quantum optics. Some consequences of the present work on the general
description of relaxation in quantum dots are given as well, and
conclusions are drawn in Section IV. In the Appendix, the coupling
constants are calculated adopting parameter values for an idealized
GaAs quantum dot with parabolic confinement.

\section{The Model and the Transformation}
%
Let us consider $N$ discrete electronic levels $i = 0 \dots N-1$ coupled to
$M$ phonon modes $\bf{q}_1 \dots \bf{q}_M$ of fixed energy $\hbar\omega_0$.
The Hamiltonian reads
\begin{eqnarray} \label{themodel0}
  H& = & \sum_{i}\epsilon_i \,c_i^{\dagger}\,c_i +
           \sum_{\mathbf{q}} \hbar\omega_0 \, b_{\mathbf{q}}^{\dagger} \,
           b_{\mathbf{q}}  \nonumber \\
  & + & \sum_{i,j,\mathbf{q}} M_{\mathbf{q}}^{i,j}
          \left(b_{\mathbf{q}} + b_{\mathbf{-q}}^{\dagger}\right)
           c_i^{\dagger}\,c_j \; .
\end{eqnarray}
Here, the $c_i(c_i^\dagger)$ denote the fermionic creation (annihilation)
operators, respectively, and $b_{\mathbf{q}}, b_{\mathbf{q}}^\dagger$
are the corresponding bosonic operators. The coupling constants between
phonons and electrons, $M_{\mathbf{q}}^{i,j}$,  depend explicitly on the
fermionic states involved (transitions between dot levels).
For $H$ to be hermitean, $(M_{\mathbf{q}}^{i,j})^*=M_{\mathbf{-q}}^{j,i}$
must hold. For simplicity, the spin is neglected because it is conserved
by the electron-phonon interaction.

Since all $M$ bosonic modes couple to the electronic levels, a straightforward
diagonalization is not feasible. We proceed by mapping the phonon operators
$\{b_{\mathbf{q}}\}$ onto a new set of bosonic operators
$\{B_{\mathbf{\lambda}}\}$, with the goal that only a limited number
couples to the electrons.

We start with an arbitrary linear combination of the operators
$b_{\mathbf{q}}$ written as $A_{\lambda}$ with $\lambda=1\dots M$,
which also span the bosonic Hilbert space.
Following the well-known Gram-Schmidt orthogonalization procedure,
\begin{equation}      \label{GramSchmidt}
\widetilde{B}_{\lambda} = A_{\lambda}-\sum_{\alpha=1}^{\lambda-1}
 [A_{\lambda},B_{\alpha}^{\dagger}] \, B_{\alpha} \; , \; \;
 B_\lambda = \widetilde{B}_{\lambda}/
\sqrt{[\widetilde{B}_{\lambda},\widetilde{B}_{\lambda}^{\dagger}]}
\end{equation}
we arrive at a new set $B_{\lambda}$ whose members obey the
canonical Bose commutation relations,
$[B_{\lambda},B_{\lambda'}]=\delta_{\lambda,\lambda'}$.
Then, the transformation matrix $U$ with
$B_\lambda = \sum_{\mathbf{q}} U_{\lambda,\mathbf{q}} b_{\mathbf{q}}$ is
unitary, and we have
\begin{equation}
\sum_{\lambda}B_{\lambda}^{\dagger}B_{\lambda}=
\sum_{\mathbf{q}}b_{\mathbf{q}}^{\dagger}b_{\mathbf{q}} \; .
\end{equation}
Since the bosonic modes have the {\it same energy}, the last relation
shows that the free boson term in the Hamiltonian Eq.(\ref{themodel0})
remains diagonal.

The first linear combinations are chosen as
\begin{equation}
A_{\lambda(i,j)}=\sum_{\mathbf{q}}M_{\mathbf{q}}^{i,j}b_{\mathbf{q}} \; ,
\end{equation}
where $\lambda(i,j)$ runs over the $N(N+1)/2$ pairs $(i,j)$ with $i\geq j$.
The remaining $A_\lambda$ can be taken arbitraryly but linearly
independent. It follows from the prescribed one-to-one mapping that the
electron-phonon interaction contains only the restricted set $\lambda(i,j)$
of the new operators. This reduces the numerical labour enormously since
the relevant Hilbert space contains now $N$ fermionic and only $N(N+1)/2$
bosonic degrees of freedom.
%
\subsection{Reduction to a Two-Level System}
%
The model can be used to describe electrons in a quantum dot which are
coupled to LO-phonons. Under the assumption that the third electronic level
is energetically well above the lowest two, we will limit the number
of states to $N=2$. Choosing
$A_1=\sum_{\mathbf{q}}M_{\mathbf{q}}^{1,0}b_{\mathbf{q}}$,
$A_2=\sum_{\mathbf{q}}M_{\mathbf{q}}^{0,0}b_{\mathbf{q}}$,
$A_3=\sum_{\mathbf{q}}M_{\mathbf{q}}^{1,1}b_{\mathbf{q}}$
we accomplish that the transition matrix element couples only to
{\em three} non-trivial bosonic modes. We obtain
\begin{eqnarray}   \label{themodel1}
& H & = \; \epsilon_0 c_0^\dagger c_0 \; + \; \epsilon_1 c_1^\dagger c_1
        \; + \; \sum_\lambda \hbar\omega_0 \,B_{\lambda}^{\dagger}B_{\lambda} \\
 & + & \left(C_1B_1+C_1^*B_1^\dagger \right)
            (c_1^{\dagger}c_0+c_0^{\dagger}c_1)  \nonumber \\
 & + & \left(C_2B_1 + C_2^*B_1^\dagger +
            C_3B_2 + C_3^*B_2^\dagger\right)c_0^{\dagger}c_0 \nonumber \\
 & + & \left(C_4B_1 + C_4^*B_1^\dagger +
            C_5B_2 + C_5^*B_2^\dagger
          +  C_6B_3+C_6^*B_3^\dagger\right)c_1^{\dagger}c_1 \nonumber
\end{eqnarray}
with the six coupling constants $C_i$ which follow from the transformation
(\ref{GramSchmidt}). If we further assume that the electronic wavefunctions
in the quantum dot exhibit a well-defined parity, the constants $C_2$
and $C_4$ vanish, and $B_1$ couples only to the transition $0 \rightarrow 1$.
The model will therefore show prominent features of the Jaynes-Cummings
model\cite{JaCum} which has been introduced to describe a single photon
mode coupled to an atomic transition. In the present case, however, the
detuning has to be defined as
$\Delta = \hbar\omega - (\epsilon_1 - \epsilon_0)$.
Further details regarding the explicit calculation of the coupling
constants $C_i$ are given in Appendix \ref{sixconstants}.
We adopt a parabolic confinement potential with extensions $y_0=z_0$ and
$x_0>y_0$, having in mind an anisotropic harmonic quantum dot.
In this case, all the lowest dot levels have equal energy separation, and
the mentioned truncation to just two levels is not realistic. Nevertheless
it will be applied here to keep the numerics at a reasonable level.

By means of an appropriate rotation of the operators $B_2$ and $B_3$, i. e.
\begin{eqnarray}
B_2 & \Leftarrow & \left( (C_5-C_3)B_2+C_6B_3\right)/\gamma \\
B_3 & \Leftarrow & \left( -C_6^*B_2+(C_5^*-C_3^*)B_3\right)/\gamma \nonumber
\end{eqnarray}
with $\gamma^2 = |C_5-C_3|^2+|C_6|^2$, we manage that the new mode $B_3$
only couples to the fermionic particle number operator
${\cal N}\equiv c_0^{\dagger}c_0 + c_1^{\dagger}c_1$.
If we leave out the bosonic modes which do not couple to the fermionic
levels at all we obtain
\begin{eqnarray}  \label{themodel2}
H & = & \epsilon_0 c_0^\dagger c_0 \; + \; \epsilon_1 c_1^\dagger c_1 \\
  & + &\hbar\omega_0B_1^{\dagger}B_1
    + (C_1B_1 + C_1^*B_1^\dagger)(c_1^{\dagger}c_0 + c_0^{\dagger}c_1)
                                       \nonumber \\
  & + &\hbar\omega_0 B_2^\dagger B_2
         + (\gamma B_2 + \gamma B_2^\dagger)c_1^{\dagger}c_1
         - (\eta^* B_2 + \eta B_2^\dagger)\mathcal{N}  \nonumber  \\
  & + &\hbar\omega_0 B_3^\dagger B_3 - (\kappa^* B_3 + \kappa B_3^\dagger)
           \mathcal{N} \nonumber
\end{eqnarray}
with the new parameters
\begin{equation}
\eta = C_3^*(C_3-C_5)/\gamma \; , \; \; \kappa = C_3^*C_6^*/\gamma \; .
\end{equation}
The Hamiltonian conserves the electron number, $[H, {\cal N}] = 0$, and
consequently the Hilbert space can be decomposed according to the
electron number (zero, one, or two). In the subspace of zero or two fermions,
the Hamiltonian can be diagonalized easily because the nondiagonal
transitions $0 \leftrightarrow 1$ are impossible here.
More demanding is the subspace of one fermion, which can, however, be
rationalized a lot by introducing shifted operators
\begin{equation}  \label{Bshifted}
{\cal B}_2 = B_2 - \eta {\cal N}+\gamma c_1^{\dagger}c_1 \; , \; \;
{\cal B}_3 = B_3 - \kappa {\cal N}
\end{equation}
Note that from Eq.(\ref{Bshifted}) onward, the LO energy $\hbar \omega_0$ is taken
as unit of energy in the remainder of this section. We want to stress that
the shifted bosonic operators ${\cal B}_2$
and ${\cal B}_3$ still obey the canonical commutation rules, but do not
commute with the fermion operators.
The Hamiltonian Eq.(\ref{themodel2}) is now represented as
\begin{eqnarray} \label{themodel3}
H & = & \epsilon_0 c_0^\dagger c_0 \; + \; \tilde{\epsilon}_1^{\cal N}
   c_1^\dagger c_1 \; - \; E_p^{\cal N} \nonumber \\
  & + & B_1^ \dagger B_1 + {\cal B}^\dagger_2 {\cal B}_2
        + {\cal B}_3^\dagger {\cal B}_ 3 \\
  & + &(C_1B_1 + C_1^*B_1^\dagger)(c_1^\dagger c_0 + c_0^\dagger c_1) \; ,
       \nonumber
\end{eqnarray}
with the modified energies depending of the number of particles
\begin{eqnarray}        \label{levelshift}
\tilde{\epsilon}_1^{{\cal N}}&=&\epsilon_1 +
2\text{Re}(\gamma\eta^*) {\cal N}-|\gamma|^2 \\
                       \label{polaronshift}
E_p^{{\cal N}}& = &(|\eta|^2+|\kappa|^2){\cal N}^2 \; .
\end{eqnarray}

\begin{figure}[t]
  \begin{center}
    \includegraphics*[width=2.5in,angle=-90]{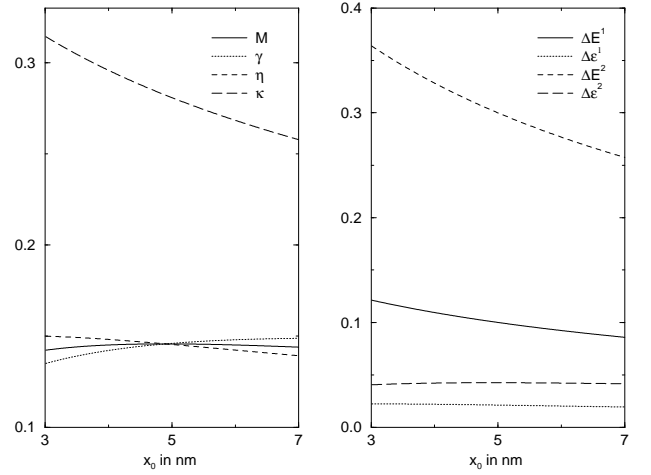}
\caption{Left: The parameters $C_1$, $\gamma$, $\eta$ and $\kappa$ in units
of the LO energy as a function of the dot size $x_0$.
Right: The relative polaronic shifts $\Delta E^N$
and the relative shifts of the level spacing $\Delta\epsilon^N$.}
\label{parameter}
\end{center}
\end{figure}
The left hand side of Fig. \ref{parameter} shows the (dimensionless)
parameters calculated for a parabolic confinement potential with
extensions of $y_0 = z_0 = 3\,$nm as a function of $x_0$. All input parameters
refer to GaAs, e.g. $\hbar\omega_0=36.7\,$meV (see Appendix \ref{sixconstants}).
The coupling constants are almost independent of the dot size and of
order $0.1\hbar\omega_0$. This value of 3.6\,meV compares favourably with
the polaron shift of electrons in bulk GaAs, 2.3\,meV.

The right hand side of Fig. \ref{parameter} shows the polaronic shift
due to the phonons $B_2$ and $B_3$, Eq.(\ref{polaronshift}),
and the renormalisation of the level spacing Eq.(\ref{levelshift}).
Since both depend on the number of electrons present, we display the
relative shifts when adding one electron, $\Delta E^N = E_p^N - E_p^{N-1}$
and $\Delta\epsilon^N = \tilde{\epsilon}_1^N - \tilde{\epsilon}_1^{N-1}$.
%

\subsection{Solution of the Model}
%
The eigenvectors $|N;n_1,n_2,n_3\rangle$ for the $N$-electron Hilbert space
with $N=0, 2$ are simple and given by
\begin{eqnarray}    \label{zeroeigenvectors}
|0;n_1,n_2,n_3\rangle & \equiv &
|V\rangle \, |n_1\rangle_0^1 \, |n_2\rangle_0^2 \, |n_3\rangle_0^3\\
|2;n_1,n_2,n_3\rangle & \equiv &
c_1^\dagger \, c_0^\dagger \, |V\rangle \, |n_1\rangle_0^1 \,
|n_2\rangle_{-2\eta+\gamma}^2 \, |n_3\rangle_{-2\kappa}^3  \nonumber
\end{eqnarray}
with the corresponding eigenvalues
\begin{equation}
E_{n_1,n_2,n_3}^N = (\epsilon_0+\tilde{\epsilon}_1^2-E_p^2)N/2
  + n_1+n_2+n_3 \; .
\end{equation}
Here, $|V\rangle$ denotes the electron vacuum, and
\begin{equation}       \label{coherentstate}
|n \rangle_{\alpha}^\lambda \equiv (n!)^{-1/2}
  (B_\lambda^\dagger + \alpha)^n |0\rangle_\alpha^\lambda
\end{equation}
are the shifted oscillator eigenstates (or coherent states) whose vacuum
is defined as $(B_\lambda + \alpha)|0\rangle_\alpha^\lambda = 0$. We
will omit the upper index $\lambda=1, 2, 3$ of the bosonic states from
now on since confusions are unlikely.

Considering the $N=1$ Hilbert space the Hamiltonian can be easily
diagonalized numerically since only {\it two} bosonic modes are involved.
An approximate analytical solution is possible as well, where the
transition matrix element is treated in degenerate perturbation theory,
using the coherent states Eq.(\ref{coherentstate}) as basis.
In oder to show the basic features of this analytic
solution we neglect for the moment the third bosonic mode ${\cal B}_3$
which reduces the number of quantum numbers that have to be kept track of.
Since this mode commutes with the rest of the system, it can easily
be incorporated afterwards. In the same spirit we neglect the polaronic
shift (\ref{polaronshift}) and the shift of the level spacing
(\ref{levelshift}) which are restored, however, in the numerics.

In the resonant situation (zero detuning $\Delta$)
and neglecting terms of the order $\gamma^2 C_1$,
{\it degenerate} pertubation theory leads to the level repulsion (or
avoided level crossing)
\begin{equation}          \label{degpert}
E_{n_1,n_2}^{\pm}=\epsilon_0 + n_1 + n_2
\pm\sqrt{n_1}\, |{}_0\langle n_2|n_2\rangle_{\gamma} \, C_1|\;,
\end{equation}
for $n_1 \geq 1$, $n_2\geq 0$.

The overlap integrals of two displaced oscillators appearing in
Eq.(\ref{degpert}) are also known as Franck-Condon factors
which were first introduced in the theory of excited molecules.
More generally they are given by
\begin{eqnarray}
{}_0\langle n|m\rangle_\gamma & = & \frac{\gamma^{n+m}}{\sqrt{n! \, m!}}
    \, e^{-\gamma^2/2} \\
& \times & \sum_{k=0}^{\text{min}(n,m)}
(-1)^{k+n}\frac{1}{k!}\frac{n!}{(n-k)!}\frac{m!}{(m-k)!}\gamma^{-2k}
       \nonumber
\end{eqnarray}
and are related to the associated Laguerre polynomials.
Since the Franck-Condon factors $\,{}_0\langle n_2|n_2\rangle_{\gamma}$ are
oscillating as functions of $n_2$ we find a complex level structure
if many bosons are present, i.e. at elevated temperatures.
Introducing a small external broadening will lead to a continuous but still
highly structured spectral function.

The energy splitting of Eq.(\ref{degpert}) can be recovered in a truncated
Hamiltonian where only the nearly resonant transitions $0 \rightarrow 1$
with phonon absorption and $1 \rightarrow 0$ with phonon emission are kept,
\begin{eqnarray}
H^{RW} & =  \epsilon_0c_0^{\dagger}c_0+\epsilon_1c_1^{\dagger}c_1
   + B_1^{\dagger}B_1 + {\cal B}_2^{\dagger}{\cal B}_2 \nonumber \\
   & + C_1B_1c_1^{\dagger}c_0+C_1^*B_1^{\dagger}c_0^{\dagger}c_1 \; .
\end{eqnarray}
In analogy to the optical equivalent in the Jaynes-Cummings model, we use
the term {\it rotating wave approximation} (RW). $H^{RW}$ can be represented
by 2x2-matrices \cite{ZiWa} in the basis
\begin{eqnarray}
|n_1,n_2\rangle^0 & \equiv &
c_0^\dagger \, |V\rangle \, |n_1\rangle_0 \, |n_2\rangle_{-\eta} \\
|n_1,n_2\rangle^1 & \equiv & c_1^\dagger \, |V\rangle \,
|n_1-1\rangle_0 \, |n_2\rangle_{-\eta+\gamma} \; .\nonumber
\end{eqnarray}
Notice that the shift of ${\cal B}_2$ depends on the electronic level, a
feature that gives rise to the Franck-Condon factors. The corresponding
eigenfunctions are
\begin{equation}    \label{oneeigenvectors}
|n_1,n_2\rangle^{\pm} = c_{n_1,n_2}^{\pm}|n_1,n_2\rangle^0\pm
c_{n_1,n_2}^{\mp}|n_1,n_2\rangle^1
\end{equation}
with the eigenvalues
\begin{equation}    \label{oneeigenvalues}
E_{n_1,n_2}^{\pm}=\frac{1}{2}\left(\epsilon_0 + \epsilon_1\right)
  + n_1 + n_2 -1/2 \pm R_{n_1,n_2}/2\;,
\end{equation}
with the Rabi splitting
$R_{n_1,n_2}^2 = \Delta^2 + n_1\,|{}_0\langle n_2|n_2
\rangle_{\gamma}\,C_1|^2$ and the weight factors
$c_{n_1,n_2}^{\pm} = [(R_{n_1,n_2}\pm\Delta)/2R_{n_1,n_2}]^{1/2}$.

In this approximation, $H^{RW}$ has been mapped onto the Jaynes-Cummings
model plus an additional bosonic mode which merely renormalizes the coupling
constant, $C_1\rightarrow {{}_0\langle n_2|n_2\rangle_{\gamma} \, C_1}$.
%
\section{The Spectral Function}
%
\label{sectionspectral}
In this section we calculate the spectral function of our system.
We contrast two different approaches and also provide the full solution
via numerical diagonalization of a matrix spanned by only two bosonic
modes.
%
\subsection{Analytic Expressions}
%
The spectral function of the electron-phonon Hamiltonian Eq.(\ref{themodel0})
is usually calculated by means of finite-temperature Green's functions
(Chapter 6 in Ref.\,\onlinecite{Mahan}). The selfconsistent first-order Born
approximation for the (retarded) selfenergy is often used provided the
assumption of weak coupling holds. With some simplification it has been
applied to the present quantum dot problem with two levels by Kr\'al
{\it et al.}\cite{KralKhas} There are two diagrams of
first order in the phonon propagator which can be classified as Hartree
and exchange selfenergy, $\Sigma = \Sigma^H + \Sigma^X$. Usually, the
Hartree term is neglected in view of a constant electron charge density
which is compensated by a positive background. However, in the case of
quantum dots being localized in space, the situation is different: In a
phonon-assisted transition between levels, the charge structure of the
electron (given by the confinement functions) changes, and a
classical electrostatic contribution to the lattice deformation appears.
Explicitly, the Hartree selfenergy of level $i$ is given by ($\hbar = 1$)
\begin{equation}        \label{Hartree}
\Sigma_i^H = -\frac{2}{\omega_0}
\sum_{j,\mathbf{q}} n_j \, M_{\mathbf{q}}^{i,i} \, M_{-\mathbf{q}}^{j,j} \; .
\end{equation}
The electronic occupations numbers $n_i$ have to be determined via the
spectral function,
\begin{equation}
n_i = \int_{-\infty}^{\infty}d\omega \, A_i^{\text{B}}(\omega)\,f(\omega)
 \; .
\end{equation}
Under equilibrium conditions, the electrons are distributed over energy
according to the Fermi function
$f(\omega)=(e^{\beta(\omega - \mu)}+1)^{-1}$ with inverse temperature
$\beta=1/k_BT$ and chemical potential $\mu$. Spectral function and
selfenergy are related as usually via
\begin{equation}
A_i^{\text{B}}(\omega) = -\frac{1}{\pi}\,\text{Im}
 \frac{1}{\omega-\epsilon_i - \Sigma_i(\omega)} \; .
\end{equation}
The exchange selfenergy reads
\begin{eqnarray} \label{selfenergy}
&\Sigma&_i^X(\omega) =
\sum_{j,\mathbf{q}}M_{\mathbf{q}}^{i,j} \, M_{-\mathbf{q}}^{j,i}
\int_{-\infty}^\infty d\omega'A_j^{\text{B}}(\omega') \\
&\times&\left[\frac{N(\omega_0) + f(\omega')}
  {\omega+\omega_0 - \omega'+i0} +
\frac{N(\omega_0) + 1 - f(\omega')}
    {\omega - \omega_0 -\omega'+i0}\right] \; . \nonumber
\end{eqnarray}
Here, $N(\omega)=(e^{\beta\omega}-1)^{-1}$ is the Bose function and
gives the phonon occupation. For the selfconsistent first-order Born
approximation, Eqs.(\ref{Hartree}-\ref{selfenergy}) have to be solved
in an iterative manner until convergency for the total selfenergy
$\Sigma_i(\omega)$ is reached.

This approximate result wil be compared with the spectral function
based on the numerically exact eigenfunctions and eigenvalues.
As starting point we use the general definition
\begin{equation}      \label{definitionspectral}
A_{i,j}(\omega) = \frac{1}{\pi} \, \text{Re}
\int_{0}^{\infty} dt \, e^{i\omega t} \,
\text{Tr} \left(\rho \, [c_i(t),c_j^{\dagger}(0)]_+ \right)
\end{equation}
with $\rho$ as the density matrix. Expanding into exact eigenstates we
obtain
\begin{eqnarray}      \label{Nspectral}
A_{i}(\omega) & =& \sum_{N,\nu,\mu}
    \left( \rho_\nu^N+\rho_\mu^{N-1} \right) |\langle
   N-1;\mu|c_i|N;\nu \rangle|^2 \nonumber \\
& & \times \, \delta(\epsilon_{\nu}^N-\epsilon_{\mu}^{N-1}-\omega) \; ,
\end{eqnarray}
where $H|N;\nu\rangle=\epsilon_{\nu}^N|N;\nu\rangle$ denotes the eigenvalue
problem in the $N$-electron subspace. In equilibrium, $\rho$ is diagonal,
$\langle N;\nu|\rho|N;\mu\rangle = \delta_{\nu,\mu} \, \rho_{\nu}^N$.
Due to the well-defined parity of the Hamiltonian
Eq.(\ref{themodel3}), the spectral function has only elements diagonal
in the sublevel index. Parity means here that the electron-$B_1$ subspace
decomposes into two subsets defined by the states
$\{c_0^\dagger|V\rangle|2n_1\rangle_0\, , \; c_1^\dagger|V\rangle
|2n_1+1\rangle_0\}$ and
$\{c_0^\dagger|V\rangle |2n_1+1\rangle_0 \, , \; c_1^\dagger
|V\rangle  |2n_1\rangle_0\}$, respectively.

For the spectral function in the rotating wave approximation, an explicit
result can be given using the results of
Eqs.(\ref{oneeigenvectors},\ref{oneeigenvalues}). Again the mode ${\cal B}_3$
is neglected which can be easily incorporated afterwards. To complete
the spectrum of $H^{RW}$ we have to include the non-interacting states
$|0,n_2\rangle^0\equiv
c_0^{\dagger} \, |V\rangle \, |0\rangle_0|n_2\rangle_{-\eta}$ with
eigenvalues $\epsilon_0+n_2\omega_0$, which resemble the groundstate
for fixed $n_2$.

The spectral function in RW has two contributions,
$A_0^{\text{RW}} = A_0^{\text{RW},10} + A_0^{\text{RW},21}$, which refer
to the transition between one/zero particles and two/one particles,
repectively.
\begin{eqnarray}  \label{ARW}
& A &_0^{\text{RW},10} = \sum_{n_2,n_2'=0}(\rho_{0,n_2}^1+\rho_{0,n_2'}^0)
    |{}_0\langle n_2'|n_2\rangle_{\eta}|^2\\\nonumber
& \times &\delta(\epsilon_0+(n_2-n_2')\omega_0-\omega)\\\nonumber
& + &\sum_{n_1=1;n_2,n_2'=0}(\rho_{n_1,n_2,-}^1+\rho_{n_1,n_2'}^0)
     |c_{n_1,n_2}^-|^2|{}_0\langle n_2'|n_2\rangle_{\eta}|^2\\\nonumber
& \times &\delta(\epsilon_0-\Delta/2-R_{n_1,n_2}/2+(n_2-n_2')\omega_0-\omega)\\
      \nonumber
& + &\sum_{n_1=1;n_2,n_2'=0}(\rho_{n_1,n_2,+}^1+\rho_{n_1,n_2'}^0)
    |c_{n_1,n_2}^+|^2|{}_0\langle n_2'|n_2\rangle_{\eta}|^2\\\nonumber
& \times &\delta(\epsilon_0-\Delta/2+R_{n_1,n_2}/2+(n_2-n_2')\omega_0-\omega)
\; ,
\end{eqnarray}
where $\rho_{0,n_2}^1\equiv{}^0\langle 0,n_2|\rho|0,n_2\rangle^0$,
$\rho_{n_1,n_2,\pm}^1\equiv{}^{\pm}\langle n_1,n_2|\rho|n_1,n_2\rangle^{\pm}$
and $\rho_{n_1,n_2}^0\equiv\langle 0;n_1,n_2|\rho|0;n_1,n_2\rangle$ [see
Eq.(\ref{zeroeigenvectors})].
An analogous expression can be given for $A_0^{\text{RW},21}(\omega)$.
The spectral function $A_0^{\text{RW}}(\omega)$ satisfies the
strict sum rules for zeroth and first moment which give normalization
and average energy, respectively.

Since the overlap between displaced oscillators does not vanish even
for different quantum numbers, the double summation over $n_2$ and $n_2'$
will account for the satellites of the spectral function which appear at
multiples of the phonon frequency $\omega_0$. The approximation is
good as long as the system is close to zero detuning, i.e.
as long as degenerate pertubation theory works. Treating additionally the
eigenstates in perturbation theory leads to marginal improvements only.

For the construction of the one-particle spectral function,
a combination of states referring to different fermion subspaces are
needed, since the spectral function gives the frequency-resolved
probability for removal (or addition) of one electron.
A signature of this general behaviour is the occurence of both the
Rabi splitting and the bare detuning in the delta function argument of
Eq.(\ref{ARW}), accompanied with the overlap between non-interacting
and interacting states. Since, on the other hand, the phonon-assisted
transition of an electron occurs exclusively in the subspace of one
fermion, we have severe doubts on using a convolution of one-particle
spectral functions for describing relaxation, as done in Ref.\cite{KralKhas}.
We expect this deficiency to be dramatic in particular when only a few
electronic levels are involved, as in the present case.
%
\subsection{Numerical Results}
%
In the following figures we present our numerical results. The exact spectral
function of the lower level $A_0(\omega)$ (solid line), obtained from
numerical diagonalization, is compared with the simplified solution in
rotating wave approximation $A_0^{\text{RW}}$ (dotted line),
and with the selfconsistent first-order Born approximation $A_0^{\text{B}}$
(dashed line). Both $A_0$ and $A_0^{\text{RW}}$ are
convoluted with a Gaussian of variance $\sigma=\hbar\omega_0/70$.

However, we want to emphasize that the exact spectral function (and
$A_i^{\text{RW}}(\omega)$ as well) consists of delta functions only.
This was clear from the beginning since a finite pertubation cannot change
the character of the spectrum of the unperturbed system. Since we started
from {\it dispersionless} bosonic modes, the discrete electronic
spectrum cannot be altered by the finite electron-phonon
interaction.\cite{ReSi}
For the density matrix in Eq.(\ref{definitionspectral}) we assume the
grand canonical equilibrium distribution
$\rho\propto\exp(-\beta (H-\mu{\cal N}))$ and fix the chemical potential
$\mu$ halfway in energy between the bare electronic levels.
This is close to putting just one electron into the dot.

\begin{figure}[t]
  \begin{center}
    \includegraphics*[width=2.5in,angle=-90]{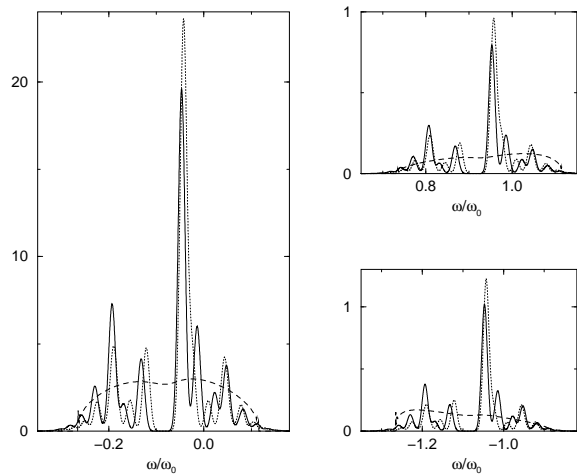}
\caption{The spectral functions $A_0$ (full), $A_0^{\text{RW}}$ (dotted)
and $A_0^{\text{B}}$ (dashed) at $k_BT=\hbar\omega_0$ and zero detuning.
The energy scale is in units of $\hbar\omega_0$, with $\epsilon_0 = 0$.
The right panels show the first satellite structure at one phonon energy
above (top) and below the main level (bottom).}
\label{E1omega}    
\end{center}
\end{figure}

Figure \ref{E1omega} shows the spectral functions at $k_BT=\hbar\omega_0$
($T = 426\,$K) for the {\it bare} detuning zero, {\it i. e.}
$\epsilon_1=\hbar\omega_0$ ($\epsilon_0$ is taken as zero of energy in what
follows). This level spacing corresponds to a dot size of $x_0=3.9\,$nm.

\begin{figure}[t]
  \begin{center}
    \includegraphics*[width=2.5in,angle=-90]{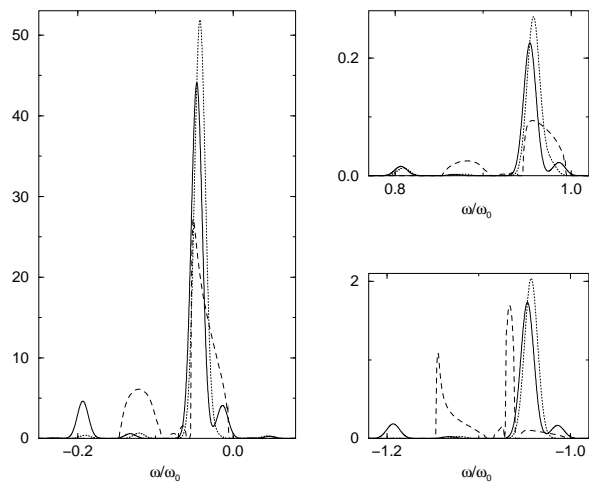}
\caption{The spectral functions at $k_BT=\hbar\omega_0/4$ and zero
detuning. Other data as in Fig. 2}	
\label{E1omega/4}
  \end{center}
\end{figure}

The prominent feature is the splitting of the spectral function into
some kind of dublett which can be traced back to the level repulsion
in the (dominant) $B_1$-channnel. Its value is related to the strength of
$C_1$ (3.3 meV). Looking more closely it becomes apparent, however, that
the upper and lower structure have not an equal weight as one would expect
for zero detuning. This clearly indicates the shift in level spacing
(renormalized detuning). Further, an overall shift with respect to
the bare energy $\epsilon_0=0$ is obvious which stems from the polaronic
shift of the other modes. All these features can be seen also at
$k_BT=\hbar\omega_0/4$ ($T = 106\,$K) which is more
relevant for realistic dot spectroscopy (Fig. \ref{E1omega/4}).
The right hand panels of Figs. \ref{E1omega} and \ref{E1omega/4}
show the first satellite structures. They are reduced in weight but have
a similar appearance as in the main structure.

\begin{figure}[t]
  \begin{center}
    \includegraphics*[width=2.5in,angle=-90]{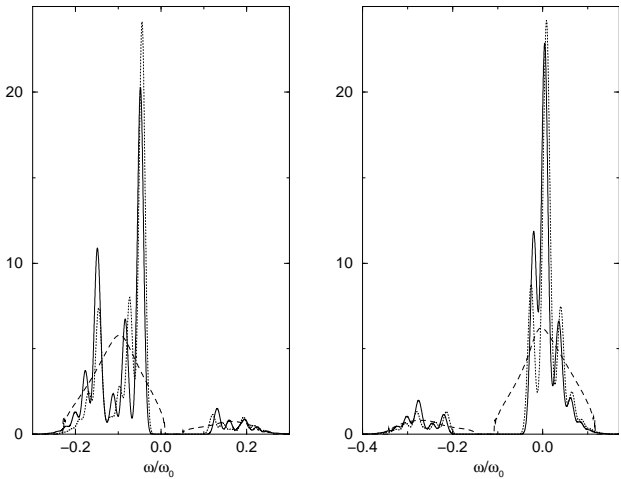}
    \caption{The spectral functions at $k_BT=\hbar\omega_0$ for negative
detuning ($\Delta = -0.2\,\hbar\omega_0$, left panel) and positive detuning
($\Delta = 0.2\,\hbar\omega_0$, right panel). Other data as in Fig. 2}
\label{E112omega08}
\end{center}
\end{figure}
In Figure \ref{E112omega08} the spectral functions at $k_BT=\hbar\omega_0$
are compared for two nonzero detunings
$\Delta = \hbar\omega_0 - \epsilon_1 + \epsilon_0$ which refer to dot
sizes of $x_0=3.6\,$nm (left) and $x_0=4.4\,$nm (right), respectively.

The spectral function $A_0^{\text{RW}}(\omega)$ shows good agreement with
the exact solution in all features. In particular, the complex structure
of lines having different weight and position is well reproduced. This
means that not only the eigenvalues are correctly approximated by the
reduction of the system into (many) 2x2-matrices, but also the eigenvectors.

Turning to the selfconsistent first-order Born approximation,
$A_0^{\text{B}}(\omega)$, the agreement is not satisfying at all. At the
higher temperature (Fig. \ref{E1omega}), instead of the complex structure
only two broad bands are seen which have nearly smeared out the level
repulsion. For the lower temperature (Fig. \ref{E1omega/4}), the Born
approximation works somewhat better since here the unperturbed but shifted
ground state energy carries the dominant weight.

%
\section{Conclusions}
%
We have presented a model that is suitable to describe the
electron-phonon interaction in quantum dots and can be treated numerically
{\it exact}. Thus we were able to compare the exact spectral function with
the selfconsistent first-order Born approximation. The first observation was
that the exact spectral function consists of delta functions whereas the
Born approximation gives a continuous spectral function.
But even after an artificially broadening of the exact spectral function,
striking differences remain which are due to the unability of the Born
approach to exhbit all shifts and splittings in detail.
We would like to mention that without including the Hartree selfenergy into
the Born calculation, the agreement would be even worse.

As an alternative we propose an analytic solution which employs the
coherent states for representing the electron-phonon Hamiltonian.
The rotating wave approximation allows to reduce the problem to an
(infinite) number of decoupled 2x2-matrices. Close to resonance
(zero detuning), this approximation works remarkably good.
The appearance of Franck-Condon factors points to the great similarity
of the electron-phonon interaction in molecules and quantum dots.

Starting from the full electron-LO-phonon Hamiltonian with Fr\"ohlich
interaction, we were able reduce the problem in a quantum dot to a few
boson modes only. Due to this considerable reduction in numerical
labour, it was possible to construct the exact eigenstates and energies.
In the present work, these results have been used to calculate the
one-particle spectral function. However, an extension to two-particle
expectation values which are relevant for e.g. transition rates is possible.
Using the density-matrix scheme, relaxation processes can be studied within
the same reduced Hilbert space, a route which will be followed in future
work.
%
%
\begin{acknowledgments}
We gratefully acknowledge the support of this work by the
Deutsche Forschungsgemeinschaft in the frame of Sfb 296.
\end{acknowledgments}
%
\begin{appendix}
%
\section{Determination of coupling constants}
\label{sixconstants}
With the transformations of Eq. (\ref{GramSchmidt}) we obtain the
following expressions for the six coupling constants:
\begin{eqnarray}
|C_1|^2&=&\sum_{\mathbf{q}}
         M_{\mathbf{q}}^{1,0}M_{\mathbf{-q}}^{0,1} \nonumber \\
     C_2&=&\sum_{\mathbf{q}}M_{\mathbf{q}}^{0,0}M_{\mathbf{-q}}^{0,1}/C_1
           \nonumber \\
     |C_3|^2&=&\sum_{\mathbf{q}}M_{\mathbf{q}}^{0,0}M_{\mathbf{-q}}^{0,0}
         -|C_2|^2  \nonumber \\
     C_4&=&\sum_{\mathbf{q}}M_{\mathbf{q}}^{1,1}M_{\mathbf{-q}}^{0,1}/C_1
          +C_5C_2/C_3 \\
     C_5&=&\sum_{\mathbf{q}}M_{\mathbf{q}}^{1,1}M_{\mathbf{-q}}^{0,0}/C_3
             \nonumber \\
     |C_6|^2&=&\sum_{\mathbf{q}}M_{\mathbf{q}}^{1,1}M_{\mathbf{-q}}^{1,1}
         -|C_4|^2-|C_5|^2 \nonumber \\
        &+&2\text{Re}\left[(2C_2^* C_3C_4C_5^* -|C_2|^2|C_5|^2)
         /C_3^2\right] \; . \nonumber
\end{eqnarray}
The standard Fr\"ohlich coupling for the electron-LO-phonon
interaction is adopted and applied to the dot confinement states,
\begin{eqnarray}      \label{Psi}
M_{\mathbf{q}}^{i,j}&=&\frac{A}{V^{1/2}q} \Phi_{\mathbf{q}}^{i,j}\;,\; \;
    A^2=\hbar\omega_0\frac{e^2}{2\epsilon_0}
        \left(\frac{1}{\kappa_{\infty}}
         -\frac{1}{\kappa_0}\right) \; , \nonumber \\
    \Phi_{\mathbf{q}}^{i,j}&=&\int d^3r\, \psi_i^*({\mathbf{r}})
      e^{i{\mathbf q}\cdot{\mathbf{r}}}\psi_j({\mathbf{r}}) \; .
\end{eqnarray}
For simplicity we consider an anisotropic parabolic potential as dot
confinement, with $x$ as the long axis. The two energetically lowest
wave functions read
\begin{eqnarray}
\psi_0(\mathbf{r})& = &\left(\sqrt{2\pi} x_0 y_0 z_0\right)^{-1/2}
  \exp-\frac{1}{4}
  \left(\frac{x^2}{x_0^2} + \frac{y^2}{y_0^2} + \frac{z^2}{z_0^2} \right)\;,
    \nonumber \\
 \psi_1(\mathbf{r})& = &\frac{x}{x_0}\psi_0(\mathbf{r}) \; ,
\end{eqnarray}
where $x_0>y_0, z_0$ are the spatial extensions (variances) of the
groundstate. The relevant level distance is
\begin{equation}    \label{Leveldist}
\epsilon_1 - \epsilon_0 = \frac{\hbar^2x_0^{-2}}{2 m_e} \; .
\end{equation}
The matrix elements in Eq.(\ref{Psi}) read
\begin{eqnarray}
\Phi_{\mathbf{q}}^{0,0}& = &\exp -\frac{1}{2}\left(q_x^2 x_0^2 +
q_y^2 y_0^2 + q_z^2 z_0^2\right)  \nonumber \\
\Phi_{\mathbf{q}}^{1,0}& = &iq_x x_0 \Phi_{\mathbf{q}}^{0,0} \\
\Phi_{\mathbf{q}}^{1,1}& =
    &\left(1 - q_x^2 x_0^2\right)\Phi_{\mathbf{q}}^{0,0} \; . \nonumber
\end{eqnarray}
The final integration over $\mathbf{q}$ leads to $C_2=0$, $C_4=0$ on
account of the well-defined parity of the wave functions. The other
constants can be reduced to the following (elliptic) integrals,
\begin{equation}     \label{In}
{I}_n = \frac{A^2}{8 \pi^{3/2} x_0}
      \int_1^{\infty}dt \, t^{-n}
         \left[t(t-\alpha_y)(t-\alpha_z)\right]^{-1/2}\;,
\end{equation}
with $\alpha_y=1-(y_0/x_0)^2<1$ and $\alpha_z=1-(z_0/x_0)^2<1$. We
obtain finally
\begin{eqnarray}
C_1& = &(I_1/2)^{1/2} \; , \; \; C_3=(I_0)^{1/2} \nonumber \\
C_5& = &(I_0-I_1/2)/C_3\\
C_6& = &(I_0-I_1+3I_2/4 - C_5^2)^{1/2} \; . \nonumber
\end{eqnarray}
For the numerical calculations we choose material constants of GaAs,
i.e.\ LO phonon energy $\hbar\omega_0=36.7\,$meV, conduction band mass
$m_e = 0.067 m_0$, and dielectric constants $\kappa_{\infty}=10.7$,
$\kappa_0=12.4$. According to Eq.(\ref{Leveldist}), a (long) dot extension of
$x_0 = 3.9$\,nm gives resonance between level spacing and LO energy.
The effective dot extensions in $y$ and $z$ direction are taken equal,
$y_0 = z_0 = 3$\,nm, which renders the elliptic integrals (\ref{In})
to be simple logarithmic functions.
\end{appendix}
%

\end{document}